\documentclass[twocolumn,aps,prd,showpacs,nofootinbib,superscriptaddress,preprintnumbers,floatfix,10pt]{revtex4-1}

\usepackage{amsmath}
\usepackage{amsfonts}
\usepackage{amssymb}
\usepackage{epsfig}
\usepackage{graphicx}
\usepackage{color}
\usepackage{slashed}
\usepackage{epstopdf}
\usepackage[utf8]{inputenc}

\usepackage{hyperref}

\newcommand{\be}{\begin{equation}}
\newcommand{\ee}{\end{equation}}
\newcommand{\bea}{\begin{eqnarray}}
\newcommand{\eea}{\end{eqnarray}}

\begin{document}

\preprint{}

\title {Flowing to the continuum in discrete tensor models for quantum gravity 
}

\author{Astrid Eichhorn}
\email[]{a.eichhorn@thphys.uni-heidelberg.de} 
\affiliation{Institute for Theoretical Physics, University of Heidelberg, Philosophenweg 16, 69120 Heidelberg, Germany}

\author{Tim Koslowski}
\email[]{koslowski@nucleares.unam.mx} 
\affiliation{Instituto de Ciencias Nucleares, Universidad Nacional Aut\'onoma de M\'exico, Cto. Exterior S/N, C.U., Del. Coyoac\'an, CDMX, M\'exico}

\begin{abstract}
\noindent  Tensor models provide a way to access the path-integral for discretized quantum gravity in $d$ dimensions. As in the case of matrix models for two-dimensional quantum gravity, the continuum limit can be related to a Renormalization Group fixed point in a setup where the tensor size $N$ serves as the Renormalization Group scale. We develop functional Renormalization Group tools for tensor models with a main focus on a rank-3 model for three-dimensional quantum gravity. We rediscover the double-scaling limit and provide an estimate for the scaling exponent. Moreover, we identify two additional fixed points with a second relevant direction in a truncation of the Renormalization Group flow. The new relevant direction might hint at the presence of additional degrees of freedom in the corresponding continuum limit.

\end{abstract}


\maketitle
\section{Introduction}
Several approaches to quantum gravity implement a discrete structure of spacetime. This is generally expected to lead to a nontrivial phase structure for gravity: One phase, a ``pre-geometric" phase, consists  of building blocks of spacetime,  which are not connected in a way that is like spacetime in today's universe. A second phase, which can be reached if the fundamental interactions in the model are tuned appropriately, corresponds to a geometric phase of spacetime, in which the discrete building blocks ``condense" to form a continuum spacetime. This phase structure can be interpreted in two conceptually very different ways: In one interpretation, the discreteness of the fundamental building blocks is physical, in which case the ``condensation" of building blocks can be understood as a physical mechanism, see, e.g., \cite{Gielen:2013kla, Oriti:2016acw},  see also \cite{Surya:2011du}. In a different interpretation, the discreteness is viewed purely as a mathematical tool, that allows one to rewrite the path-integral for continuum quantum gravity in a discrete fashion. This provides a basis for the application of  
Monte Carlo algorithms to evaluate the discrete path integral, as, e.g., in Causal Dynamical Triangulations,  see \cite{Ambjorn:1998xu,Ambjorn:2011cg,Ambjorn:2012jv, Ambjorn:2012ij,Ambjorn:2016cpa,Ambjorn:2016mnn}. The physical content of the theory only emerges, if the continuum limit is taken. This is akin to lattice formulations of quantum field theory, where a lattice regularization is introduced to ensure the existence of a family of partition functions, but the  physical content of the model is exclusively contained in the continuum limit.

A specific example of such models are matrix models for two dimensional quantum gravity \cite{Weingarten:1982mg, David:1984tx, David:1985nj,Ambjorn:1985az,Kazakov:1985ea,Boulatov:1986mm,Boulatov:1986jd}, and their generalization, tensor models \cite{Sasakura:1990fs,Gross:1990du,Godfrey:1990dt,Ambjorn:1990ge},  for reviews see, e.g., \cite{Rivasseau:2011hm,Rivasseau:2012yp,Rivasseau:2013uca,Rivasseau:2016zco,Rivasseau:2016wvy,Gurau:2016cjo}. There, the path integral over random geometries is given by
\be
Z[J] = \int \mathcal{D}T\, e^{-S[T]+ J \cdot T},
\ee
where $T$ is a rank $ d$ tensor of size $N^{ d}$. $S[T]$ is a suitable action. For instance, for the case of matrix models, where ${ d}=2$, the action is built out of trace invariants, i.e., traces of products of matrices. This choice of action ensures that the Feynman diagrams of the matrix model admit a dual description that corresponds to all possible tesselations of two-dimensional surfaces. This allows one to interpret the matrix model action geometrically. The continuum limit in the geometric description is attained when the matrix size $N$ is taken to infinity. This limit is dominated by planar graphs, corresponding to tesselations of the sphere. To obtain contributions from higher orders in the $1/N$ expansion, one has to consider the double-scaling limit \cite{Douglas:1989ve,Brezin:1990rb,Gross:1989vs}, where $N^{5/4} (g_{\rm crit} - g)$ is held fixed, where $g_{\rm crit}$ is the critical value of the coupling. In that limit, all topologies contribute. As a generalization to $d>2$ one considers rank-$d$-tensor models  and constructs the analogous dual geometric interpretation of the tensor model Feynman graphs. For some time, a major obstacle to generalize the success-story of the two dimensional case has been the lack of a $1/N$ expansion. This changed with a breakthrough by Gurau who introduced so-called colored models, where the tensors in an invariant interaction term are distinguished by different colors \cite{Gurau:2009tw,Gurau:2010nd,Gurau:2010ba,Gurau:2011aq,Gurau:2011xq}, for a review, see \cite{Gurau:2011xp}. These models admit a well-defined $1/N$ expansion. Subsequently, it was shown that all but one of the tensors can be integrated out, yielding a model of an uncolored tensor  \cite{Bonzom:2012hw}. The interaction structure of that tensor is such that all interactions can be represented in terms of colored graphs, i.e., the indices of the tensors are distinct. These are the models that we will focus on: They transform under a $U(N) \otimes ... \otimes U(N)$ symmetry, such that each index transforms under one of the symmetry groups only. This provides the basis for a combinatorics that admits a $1/N$ expansion. The symmetry entails a distinction of the indices that can be encoded in a coloring of the corresponding strands in a graphical representation of the interaction terms.

This class of models has recently also been studied in the context of the SYK model \cite{Sachdev:1992fk}. There the tensor models were used to explore the large $N$ limit of the SYK model, which is a  model of fermions with random couplings \cite{Witten:2016iux,Klebanov:2016xxf}. There, the $\epsilon$ expansion was used to study these models. It has been shown in several other cases, that the $\epsilon$ expansion and the functional Renormalization Group can go hand in hand, see, e.g., \cite{Eichhorn:2016hdi}, with the former providing an accurate benchmark for the latter,  where the latter, once its accuracy is investigated by comparing with the benchmark results, can then also be used to explore a possible nonperturbative regime. 
 While we will focus on quantum gravity when discussing conceptual aspects of our results, our technique can directly be applied to a tensor model in a different physical context, such as, e.g., the SYK model.
 
In this paper we will develop the functional Renormalization Group approach that we previously applied to matrix models \cite{Eichhorn:2013isa,Eichhorn:2014xaa} to colored tensor models, by applying it to the pure bipartite colored\footnote{Note that there are two meanings of the word ,,colored'' that are used in group field theory and general tensor models: The first refers to models with colored tensors and global color rotation symmetry, the second refers to a tensor model with colored indices. Here we use the second meaning.} rank- 3 tensor model. We use the functional Renormalization Group to investigate fixed points of the model as the tensor size $N$ goes to infinity. The dual geometric interpretation of the Feynman graphs of the model provides an interpretation of the large $N$-behavior of the tensor model as a continuum limit in the dual geometric description. This allows us in particular to investigate the double scaling limit of the tensor models and reproduce the benchmark result of \cite{Bonzom:2014oua} within the context of the functional Renormalization Group approach.

\section{The model}
We consider a rank-3-tensor model of a complex tensor $T$ and its complex conjugate $\bar{T}$ with a $U(N)\otimes U(N)\otimes U(N)$ symmetry, where each tensor index transforms separately, i.e., the order of the indices matters, and the $i$th index of one tensor can only be contracted with the $i$th index of a complex conjugate tensor in order for the symmetry to remain intact. An easy way to keep track of that is to assign a color to each of the indices, so that the graphical representation of tensor invariants that enter the action is provided by colored graphs with a white vertex for a tensor $T$ and a black vertex for its complex conjugate $\bar{T}$, and lines
come in three different colors, e.g., red (for the first index), green (for the second index) and blue (for the third index).

 The symmetries that we assume do not allow any index-dependent interactions, thus the theory space of our model consists of all index-dependent fully contracted expressions with the same number of $T$'s as $\bar{T}$'s where the contraction respects the order of the indices, see also \cite{Rivasseau:2014ima}.
 This is a clear distinction of the model that we investigate in this paper and group field theories, where the group Laplacian is usually used to define the kinetic term. This implies in particular that there exists a unique quadratic term in our case, which takes the form:

\be
S_{\rm kin} = T_{ijk}\bar{T}_{ijk}.
\ee

\subsection{Operators and their geometric interpretation}
 The geometric interpretation of rank 3 tensors is straightforward: To each tensor $T$ we associate a triangle with positive orientation, i.e., the colors of the edges of the triangles are ordered clockwise.  The complex conjugate tensors $\bar T$ is linked to triangles of negative orientation, i.e., the colors of the edges of the triangle are ordered counterclockwise.  The contraction of an index of color $i$ is geometrically represented by a gluing of the two edges. In this way one can associate a triangulation of a closed 2-surface with any colored bipartite tensor invariant.

The Feynman graphs of a tensor model can be interpreted as rank 4 tensor invariants by simply associating an additional color "0" to each tensor, which represents a fiducial index, whose contraction represents that a propagator connects the two vertices.  With this identification one proceeds analogous to the geometric interpretation of the rank 3 model, but now uses a geometric interpretation in which one associates a tetrahedron to each of the fiducial rank 4 tensors and the gluing of a boundary triangle  to the contraction of the fiducial indices. In this way one obtains a geometric interpretation of the Feynman diagrams as 3 dimensional simplicial complexes whose boundary  is given by the geometric interpretation of the 2 dimensional contraction patterns of the legs of the Feynman diagram.

The geometric interpretation of the Feynman graphs can be used to obtain a geometric interpretation of the action. This is done by assuming that all tetrahedra are of equal size and equilateral. One then considers the logarithm of the Feynman amplitude associated with each Feynman graph and expresses it in terms of Regge-type curvature invariants. In this way one can interpret the partition function as a sum over simplicial geometries with a geometric Boltzmann weight for each simplicial geometry.  
\subsection{Melonic operators}
An important class of interactions are so-called "cyclic melons," which are operators of the form
\be
T_{i_1j_1k_1}\bar{T}_{i_1j_1k_1}T_{i_2j_2k_1}\bar{T}_{i_2j_2k_2}...T_{i_nj_nk_{n-1}}\bar{T}_{i_nj_nk_1}.
\ee
Geometrically, this glues together neighboring triangles along two edges and then glues the remaining edges to each of the triangles's other neighbors. These operators triangulate  
 the 2-d surface of a sphere in 3 dimensions
in a manner like the surface of a melon is usually sliced. Starting at order $(T\bar{T})^3$, one can also form  non-cyclic melonic operators which do not feature the maximum number of ``submelons" (this is to say a melon slice in the picture we used above), such as $T_{ijk}\bar{T}_{ijl}T_{mnl}\bar{T}_{mok}T_{poq}\bar{T}_{pnq}$. As long as they contain at least one ``submelon", such as $T_{ijk}\bar{T}_{ijl}$ in the above case, they still triangulate a 3-sphere, as one can convince oneself by drawing the corresponding triangles.

At each order $n$ in the fields (note that $n$ must be even), one can form disconnected operators with up to $n/2$ disconnected parts. For the melonic case, these correspond to triangulations of several disconnected spheres. At a first glance, one might expect that the couplings of those operators simply have to vanish at a fixed point corresponding to a physically meaningful continuum limit. As we will see below, that expectation is fulfilled for the fixed point corresponding to the double-scaling limit. On the other hand, we will see that they approach finite fixed-point values at another fixed point that we discover. In that case, they might signal the presence of additional, non-geometric degrees of freedom.

\subsection{Notation}

 To distinguish the couplings 
 of the tensor invariants, we introduce the following notation
\be
{\bar{g}}_{i,j}^{k,m,...}
\ee
 for the coupling in front of an operator with $i$ tensors, i.e., $i/2$ $T$'s and $i/2$ $\bar{T}$'s. The number of connected components is denoted by $j$, i.e., in a ``single-trace" operator with $j=1$ one can follow a closed line (disregarding the color) from any $T$ to any $\bar{T}$. So far, this structure is reminiscent of matrix models, where matrix invariants are distinguished by the number of matrices and the number of connected components, i.e., single-trace versus multi-trace operators. The tensor-specific structure is encoded in the upper indices. Here, we will stick to a notation where the first index counts the number of ``sub-melons", i.e., pairs of the form $T_{abc}\bar{T}_{dbc}$, or $T_{abc}\bar{T}_{adc}$ or $T_{abc}\bar{T}_{abd}$. Graphically, these appear as pairs of a tensor $T$ and a complex conjugate $\bar{T}$ with \emph{two} connected lines. Further, the 2nd index (and for operators beyond $i=8$ also additional ones) denotes the distinguished color. For instance, all cyclic melons distinguish one color, which is the one that labels all single lines (and consequently does not appear as an internal line on the sub-melons).

\begin{figure}
\includegraphics[width=\linewidth, clip=true, trim=3cm 1cm 0cm 9cm]{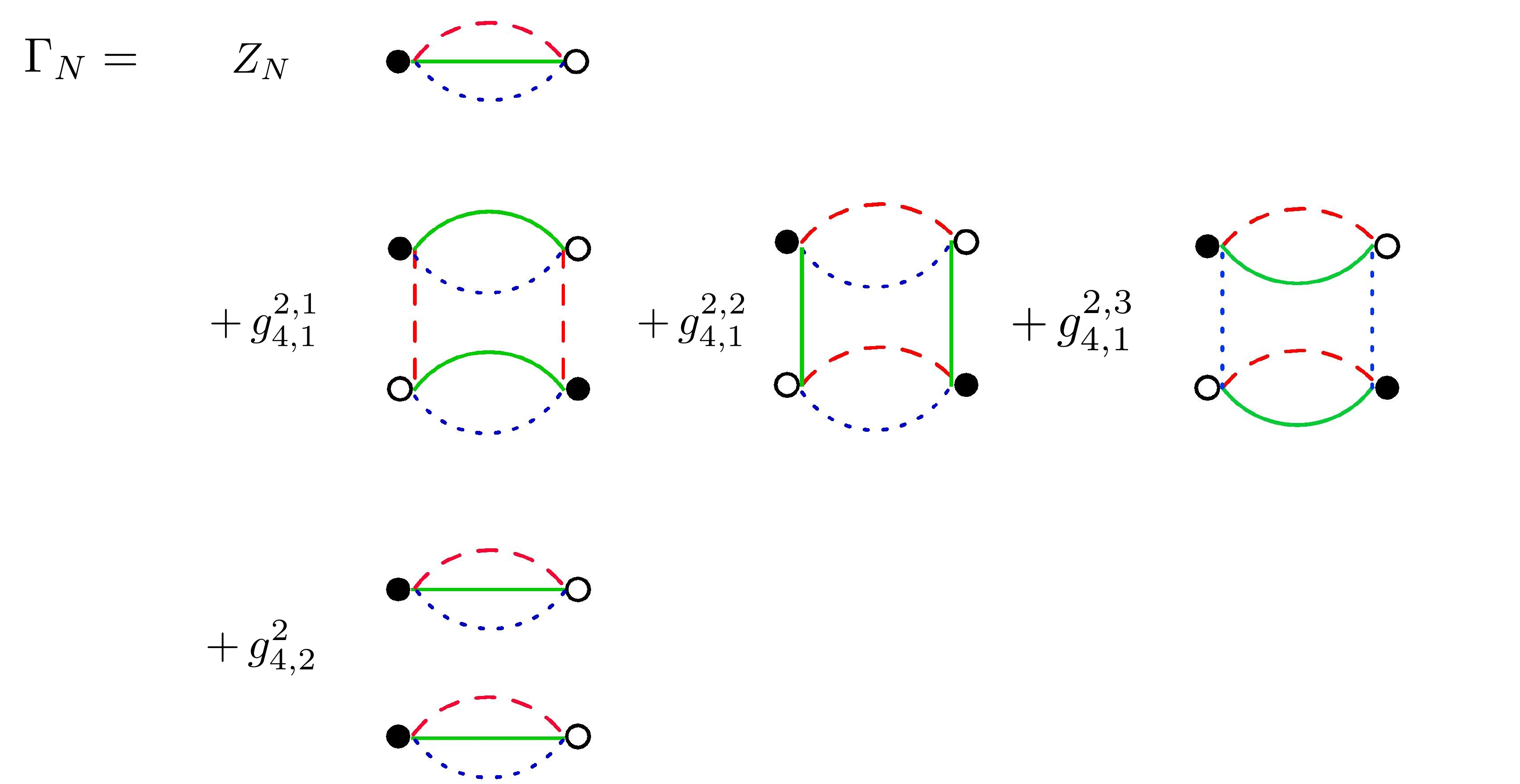}
\caption{\label{fig:simplesttrunc}We denote a tensor $T$ by a white circle and a tensor $\bar{T}$ by a black, filled circle. The first index is denoted by a red, dashed, the second by a green thick and the third by a blue dotted line.}
\end{figure}

For instance, the action containing all quadratic and quartic operators is of the form
\bea
S &=& Z_N T_{abc} \bar{T}_{abc} + {{\bar g}}_{4,1}^{2,1} \,T_{abc}\bar{T}_{dbc}T_{def}\bar{T}_{aef} \nonumber\\
&{}&+{\bar g}_{4,1}^{2,2} \,T_{abc}\bar{T}_{adc}T_{edf}\bar{T}_{eaf} \nonumber\\
&{}&+{ \bar g}_{4,1}^{2,3} \,T_{abc}\bar{T}_{abd}T_{efd}\bar{T}_{efa} \nonumber\\
&{}& + { \bar g}_{4,2}^{2}\, T_{abc} \bar{T}_{abc}\, T_{def}\bar{T}_{def}. \label{eq:simplesttrunc}
\eea
Graphically, it is represented in Fig.~\ref{fig:simplesttrunc}.

 \section{Functional Renormalization Group for tensor models}
 The FRG approach to tensor models follows the conceptual insight of \cite{Brezin:1992yc}, which suggested that a coarse-graining procedure for matrix models can be set up by using the matrix-size $N$ as the RG scale, and successively integrating out rows and columns. This idea underlies the FRG approach to matrix models, developed in \cite{Eichhorn:2013isa, Eichhorn:2014xaa}, that has subsequently been adapted to the group-field theory case \cite{Benedetti:2014qsa,Benedetti:2015yaa,Geloun:2015qfa,Geloun:2016qyb,Lahoche:2016xiq,Carrozza:2016tih,Geloun:2016xep}. Using a variant of the Polchinski equation for tensor models is also possible, see \cite{Krajewski:2015clk,Krajewski:2016svb}.
 Conceptually, this form of coarse graining, which does not rely on a local coarse-graining procedure \emph{in spacetime}, but instead makes use of a more abstract notion of coarse graining, is well-suited to quantum gravity, where local coarse-graining in spacetime is difficult to reconcile with background-independence, see, e.g., \cite{Becker:2014qya,Dietz:2015owa,Labus:2016lkh,Morris:2016spn}.  Here, we
 implement an RG flow in the sense of an interpolation between models with a large number of degrees of freedom and one with a smaller number of degrees of freedom, which describes the same physics for all coarse enough observables. This interpolation is obtained by embedding the coarser model into the finer model and integrating out the additional degrees of freedom that appear in the finer model.
 
 Note that in group field theories, for reviews see \cite{Freidel:2005qe,Oriti:2007qd,Oriti:2011jm,Carrozza:2016vsq}, which feature the combinatorical structure of tensor models, combined with a non-local quantum field theoretic structure as they are quantum field theories living on an abstract group manifold, the Renormalization Group can be set up in a more standard way, and provides many examples for perturbatively renormalizable and even asymptotically free models \cite{BenGeloun:2011rc,BenGeloun:2012pu,BenGeloun:2012yk,Carrozza:2014rya,Rivasseau:2015ova}, for a review see \cite{Carrozza:2016vsq}.
 
 We will set up a Wilsonian RG flow for the effective average action, which is defined by a modified Legendre transform based on the generating functional
\be
Z_N[J,\bar{J}] = \int \mathcal{D}T\, e^{-S[T] - T_{abc} R_N((a+b+c)/N)\bar{T}_{abc} + J \cdot T + \bar{J}\cdot\bar{T}},
\ee
where $J$ is a source and the dot is shorthand for a full contraction of the indices. $R_N((a+b+c)/N)$ is an infrared cutoff operator: As $N$ decreases, more modes are integrated out in the path integral. We define the effective average or flowing action as 
\bea
\Gamma_N[T, \bar{T}] &=& \sup_J \left(J\cdot T + \bar{J}\cdot \bar{T} - \ln Z_N[J, \bar{J}] \right) \nonumber\\
&{}&- T_{abc} R_N((a+b+c)/N)\bar{T}_{abc}. 
\eea
Note that herein $T$ denotes the expectation value of the variable in the path integral. In a slight abuse of notation we will not distinguish these here.  
We follow the steps  laid out in \cite{Eichhorn:2013isa, Eichhorn:2014xaa} to obtain a flow equation for rank-3-tensor models, of the usual structure of the one-loop equation for the flowing action \cite{Wetterich:1992yh,Morris:1993qb},  for reviews in the usual case of a continuum QFT see, e.g., \cite{Berges:2000ew, Polonyi:2001se,
Pawlowski:2005xe, Gies:2006wv, Delamotte:2007pf, Rosten:2010vm, Braun:2011pp}
\be
\partial_t \Gamma_N = {\rm Tr} \left(\Gamma_N^{(2)}+R_N \right)^{-1}\partial_t R_N,
\ee
 where $\partial_t = N \partial_N$, and
 \be
 \Gamma_N^{(2)} = \Gamma_{N\, ijk\, lmn}^{(2)}= \frac{\delta}{\delta T_{ijk}}\frac{\delta}{\delta \bar{T}_{lmn}}\Gamma_N.
 \ee
 As $N$ is changed, quantum fluctuations are integrated out in a scale-dependent fashion.
 In that process, all interactions compatible with the symmetries of the model are generated, as usual in a Wilsonian setting.  The Wetterich equation thus defines a vector field in theory space, i.e. the (infinite dimensional) space of all couplings.  This vector field encodes the scale dependence of the couplings, i.e., it provides the beta functions.

Specifically, 
 we will then employ the $\mathcal{P}^{-1}\mathcal{F}$ expansion, where 
 \bea
 \mathcal{P}&=& \Gamma_k^{(2)}\Big|_{T=0, \bar{T}=0}+R_k, \\
 \mathcal{F}&=& \Gamma_k^{(2)}-  \Gamma_k^{(2)}\Big|_{T=0, \bar{T}=0},
 \eea
 i.e., $\mathcal{F}$ contains only the field-dependent part of the inverse propagator. The flow equation can then be expanded as follows
 \bea
 \partial_t \Gamma_N = {\rm Tr} \widetilde{\partial}_t \mathcal{P}^{-1}+ \sum_{n=0}^{\infty} \frac{(-1)^{n-1}}{n}\widetilde{\partial_t} \left(\mathcal{P}^{-1}\mathcal{F}\right)^n,
 \eea
 where $\tilde{\partial}_t$ only acts on the scale-dependence of the regulator and not on the scale-dependence within $\mathcal{F}$. This provides a straightforward way of extracting the beta functions associated to tensor invariants at any given order in the tensors.
 
 \subsection{Choice of regulator}
The FRG setup for matrix models can straightforwardly be generalized to tensor models - essentially by adding the third index. Accordingly, a suitable infrared regulator takes the form
\be
R_N(i,j,k) =Z_N\left( \frac{N}{i+j+k}-1\right)\theta\left(N-(i+j+k)\right).
\ee
 This regulator term gives a mass term of order $N$ to tensor components with $i+j+k\ll N$ while not changing the kinetic term for tensor components with $i+j+k>N$, so the "index position" $i+j+k$ plays a role analogous to the total momentum in the application of the FRG in standard Euclidean field theory.

 Analogous to the case of matrix models, imposing an infrared cutoff on the matrix size breaks the  $U(N)\otimes U(N)\otimes U(N)$-symmetry of the tensor model. In particular, the fact that the cutoff should diverge as $N \rightarrow \infty$, combined with the requirement that it has the same canonical dimensionality as the kinetic term, implies that it must be a non-trivial function of the indices. Thus, this choice of cutoff cannot even preserve a  $U(N')\otimes U(N') \otimes U(N')$ subgroup of the symmetry. The only symmetry that is preserved is one a $U(1)$ that acts in the same way on all components of the tensor, i.e., it maps $T_{ijk} \rightarrow e^{i \alpha}T_{ijk}$, $\bar{T}_{ijk}\rightarrow e^{-i \alpha}\bar{T}_{ijk}$. This ensures that all interactions must feature the same number of $T$'s as of $\bar{T}$'s.

 The summation over the indices on the right-hand-side of the Wetterich equation can be performed once we have projected onto a particular operator, as we will discuss below. The sum can then be rewritten  as an integral, which takes the same value as the sum at leading order in $N$.
 \subsection{Criteria to define a truncation}
  We aim at discovering interacting fixed points underlying a possible continuum limit and characterizing their spectrum of eigenperturbations, with a particular focus on the relevant directions. 
 At an interacting fixed point, one cannot a priori know which operators are relevant, as quantum fluctuations result in corrections to canonical scaling which a priori might be large. Nevertheless, canonical dimensionality can provide a powerful guiding principle to set up truncations in the nonperturbative regime, as shown by examples in continuum quantum gravity, see, e.g., \cite{Falls:2013bv,Falls:2014tra,Gies:2016con}, as well as in the case with matter, see, e.g., \cite{Narain:2009fy, Eichhorn:2016esv,Eichhorn:2016vvy}:  There, one assumes that the effect of quantum fluctuations on the scaling dimensions is to add a finite shift of $\mathcal{O}(1)$. Then, canonically marginal or just irrelevant operators might become relevant, but operators which are highly irrelevant remain irrelevant. One can then test the consistency of this assumption by constructing a truncation according to that guiding principle and testing whether the operators do indeed follow that pattern at an interacting fixed point. In many cases,
 this principle is also supported by the fact that the fixed point can be traced to a free fixed point as one approaches the critical dimensionality of the model. For instance, the Wilson-Fisher fixed point emerges from the Gau\ss{}ian fixed point for $d<4$, and is interacting in $d=3$. There, the critical exponents follow canonical scaling, as the mass operators generates a relevant interaction, but all further interactions are irrelevant.
 
 Following this reasoning, we will use the canonical dimensionality as a guiding principle to set up truncations for tensor models. As  in the case of pure matrix models a new challenge appears in these pre-geometric models: the canonical dimensionality does not follow from straightforward scaling arguments as in the continuum case. While interactions with a higher number of tensors are increasingly irrelevant canonically, the detailed structure of the interaction plays a role in determining the scaling dimensionality. We will determine it  from the functional Renormalization Group equation directly.
 
Note that as an important check on the consistency of this procedure it is necessary to increase the truncation until the addition of further operators does not result in additional relevant directions at the fixed point of interest.

\subsection{Canonical dimensionality}
The canonical dimensionality can be determined as follows:
If all couplings $ \bar{\mathcal{G}}_i$ are expressed in terms of their dimensionless counterparts $\mathcal{G}_i = N^{-d_{\bar{\mathcal{G}}_i}}\bar{\mathcal{G}}_i$, then the leading-order term in a $1/N$ expansion of the flow equation must be dimensionless, i.e., it must be finite in the limit $N \rightarrow \infty$ (it can also be zero). This allows to determine a unique scaling dimensionality of the couplings by following an iterative procedure:
\begin{enumerate}
\item  As a normalization condition we demand that the prefactor of the kinetic term is dimensionless. As there is a one-vertex diagram proportional to $\bar g_{4,1}^{2,i}$ for each $i$, this determines (a lower limit on) the canonical dimensionality of $\bar g_{4,1}^{2,i}$. Moreover, there is a one-vertex diagram proportional to $\bar g_{4,2}^{2}$, which accordingly provides (a lower limit on) the canonical dimensionality of that coupling.
\item The consistency of the thus determined $d_{ \bar g_{4,1}^{2,i}}$ and $d_{\bar g_{4,2}^{2}}$ can be checked by evaluating the two-vertex diagrams containing those couplings. A diagram which generates a contribution to the beta function of a particular coupling without being linear in that coupling itself provides an upper limit on the dimensionality of the coupling. Together with the first step, this provides a unique assignment of dimensionality for the quartic coupling.
\item All $\bar g_{6,.}^{.,.}$ couplings are generated from three-vertex-diagrams containing $\bar g_{4,i}^{i,i}$. These determine the upper bound on the canonical dimensionality of $\bar g_{6,.}^{.,.}$.
\item The consistency of the assignment of $d_{\bar g_{6,.}^{.,.}}$ can be determined by evaluating all diagrams $\sim {\bar g}_{4,.}^{.,.}, g_{6,.}^{.,.}$ as well as the ${ \bar g}_6$-tadpole diagrams, which provide lower bounds on the canonical dimensionality.
 \end{enumerate}
 This procedure can then be iterated to higher-order truncations. Note that it fails to give a unique assignment of dimensionality, if a coupling does not appear in the beta-functions of other couplings, or if its beta function does not feature contributions from other couplings.  This is a problem for "too small" truncations, where this may happen due to the absence of effective operators that link the coupling to those whose canonical dimensionality has already been determined. We will see a specific example within our largest truncation, which features only one coupling representing a distinct combinatorial structure. Due to its being the only representative of that combinatorial structure in our truncation, there is a partial decoupling of that coupling from the RG flow which does not provide us with a unique canonical dimension for that coupling. We expect this to change in even more extended truncations.
 
The above procedure provides us with an assignment for the melonic operators with the maximum number of ``submelons":
 \be
 d_{\bar{g}_{i,j}^{i/2,.}}= -(i-2) -(j-1).
 \ee
 Thus, the couplings of highest canonical dimensionality are $g_{4,1}^{2,i}$ with dimensionality -2, $g_{4,2}^{2}$ with dimensionality -3, and $g_{6,1}^{2,i}$ with dimensionality -4. Note that the canonical irrelevance of all couplings does not preclude the existence of an interacting fixed point with relevant directions.
 
 Using this notion of canonical dimension, we introduce dimensionless couplings. In that step, we also redefine the quadratic term to have a canonical prefactor  $Z_N$, absorbing the resulting factors of $Z_N$ in the couplings:
\be
g_{i,j}^{.,.} = \bar{g}_{i,j}^{.,.}\, Z_N^{-i/2} N^{- d_{\bar{g}_{i,j}^{.,.}}}.\label{eq:dimlesscoup}
\ee

 \subsection{Projection prescription}
 Leaving the breaking of the three $U(N)$ symmetries by the regulator aside, the flowing action contains all operators that can be constructed from contractions of $i$ tensors $T$ and $i$ tensors $\bar{T}$, such that the $j$th index of a $T$ is connected to the $j$th index of some $\bar{T}$. To project uniquely onto an operator $\mathcal{O}_n$ with $n$ pairs of $T, \bar{T}$, we use the following prescription:
 As a first step, we evaluate the derivative 
 \bea
&{}& \left(\frac{\delta}{\delta T_{a_{1}b_{1}c_{1}}}\frac{\delta}{\delta\bar{T}_{d_{1}e_{1}f_{1}}}...\frac{\delta}{\delta T_{a_{n}b_{n}c_{n}}}\frac{\delta}{\delta\bar{T}_{d_{n}e_{n}f_{n}}} \mathcal{O}_n\right)\Big|_{T,\bar{T}=0}\nonumber\\
 &{}&=:\pi_{\mathcal{O}\, a_1, ...,e_1,...a_n,...,e_n}
 \eea
  In the tensor $\pi_{\mathcal{O}\, a_1, ...,e_1,...a_n,...,e_n}$, we introduce a UV cutoff on the indices, which are all restricted to be $\leq N$.
Then, the contraction 
 \be
 \pi_{\mathcal{O}\, a_1, ...,e_1,...a_n,...,e_n}\pi_{\mathcal{O}\, a_1, ...,e_1,...a_n,...,e_n}= \#\, N^{3n},
 \ee
 where the sum over each index runs from 0 to $N$ provides the number $\#$.
 We now have all ingredients required to define a projection operator for the invariant $\mathcal{O}_n$: It is given by
 \bea
 \Pi_{\mathcal{O}}&=& \frac{1}{\#\, N^{3n}}\pi_{\mathcal{O}\, a_1, ...,e_1,...a_n,...,e_n}\cdot \label{eq:projection}\\
 &{}& \cdot\frac{\delta}{\delta T_{a_{1}b_{1}c_{1}}}\frac{\delta}{\delta\bar{T}_{d_{1}e_{1}f_{1}}}...\frac{\delta}{\delta T_{a_{n}b_{n}c_{n}}}\frac{\delta}{\delta\bar{T}_{d_{n}e_{n}f_{n}}}\vert_{T=0=\bar{T}}, \nonumber
 \eea
 where it is understood that the tensor and its conjugate are only set to zero after the appropriate number of derivatives of the object of interest (typically the right-hand-side of the Wetterich equation) has been taken.

We have that $\Pi_{\mathcal{O}}\mathcal{O}=1$. Note that this only works, because we imposed the additional UV cutoff on the indices, effectively working with $N\times N \times N$ tensors. Intuitively, this resembles a restriction of the amplitude of a continuum field to be less than the IR momentum cutoff. If we had not imposed a finite UV cutoff on the size of the tensors, \eqref{eq:projection} would require taking the limit $N\rightarrow \infty$ carefully.

In contrast, the projector  vanishes at leading order in $N$ when applied to any other invariant: Clearly, it vanishes when applied to an invariant which has more or less than $n$ tensors $T$. When applied to an invariant with the same number of tensors, the derivatives with respect to $T$ and $\bar{T}$ produce a \emph{different} pattern of $\delta$'s, which follows the different way in which colored lines are drawn in the different invariants. At leading order in $1/N$, the contraction of $\pi_{\mathcal{O}\, a_1, ...,e_1,...a_n,...,e_n}$ with that structure will be supressed.

For the simplest example, let us show that this prescription distinguishes the three different cylic, melonic, connected graphs with couplings $g_{4,1}^{2,i}$, i.e., let us choose
\bea
\Gamma_{k\, 4,1}&=&  { \bar g}_{4,1}^{2,1} \,T_{abc}\bar{T}_{dbc}T_{def}\bar{T}_{aef} \nonumber\\
&{}&+{ \bar g}_{4,1}^{2,2} \,T_{abc}\bar{T}_{adc}T_{edf}\bar{T}_{eaf} \nonumber\\
&{}&+{ \bar g}_{4,1}^{2,3} \,T_{abc}\bar{T}_{abd}T_{efd}\bar{T}_{efa}.
\eea
\bea
&{}&\frac{\delta}{\delta T_{a_1b_1c_1}}\frac{\delta}{\delta \bar{T}_{d_1e_1f_1}} \frac{\delta}{\delta T_{a_2b_2c_2}}\frac{\delta}{\delta \bar{T}_{d_2e_2f_2}}  \Gamma_{k\, 4,1}\nonumber\\
&{}& =2\, { \bar g}_{4,1}^{2,1}\Bigl(\delta_{a_1,d_1}\delta_{b_1,e_2}\delta_{c_1,f_2}\delta_{b_2, e_1}\delta_{c_2,f_1}\delta_{a_2,d_2}\nonumber\\
&{}& \quad \quad\quad+ \delta_{a_1,d_2}\delta_{b_1,e_1}\delta_{c_1,f_1}\delta_{a_2,d_1}\delta_{b_c,e_2}\delta_{c_2,f_2} \Bigr)\nonumber\\
&{}& + 2\,{ \bar g}_{4,1}^{2,2} \Bigl(\delta_{a_1,d_1}\delta_{c_1,f_1}\delta_{b_1,e_2}\delta_{b_2,d_1}\delta_{a_2,d_2}\delta_{c_2,f_2}\nonumber\\
&{}& \quad \quad \quad + \delta_{a_1,d_2}\delta_{c_1,f_2}\delta_{b_1,e_1}\delta_{a_2,d_1}\delta_{c_2,f_1}\delta_{b_2,e_2} \Bigr)\nonumber\\
&{}& +2\,{ \bar g}_{4,1}^{2,3} \Bigl(\delta_{a_1,d_1}\delta_{b_1,e_1}\delta_{c_1,f_2}\delta_{c_2,f_1}\delta_{a_2,e_2}\delta_{b_2,e_2}\nonumber\\
&{}& \quad \quad \quad\delta_{a_1,d_2}\delta_{b_1,e_2}\delta_{c_1,f_1}\delta_{a_2,d_1}\delta_{b_2,d_1}\delta_{c_2,f_2} \Bigr).
\eea
We thus define
\bea
\Pi_{\mathcal{O}_{4,1}^{2,1}} &=& \frac{1}{4 N^6}  \Bigl(\delta_{a_1,d_1}\delta_{b_1,e_2}\delta_{c_1,f_2}\delta_{e_1,b_2}\delta_{f_1,c_2}\delta_{a_2,d_2}\nonumber\\
&{}& \quad \quad\quad+ \delta_{a_1,d_2}\delta_{b_1,e_1}\delta_{c_1,f_1}\delta_{a_2,d_1}\delta_{b_c,e_2}\delta_{c_2,f_2} \Bigr)\cdot \nonumber\\
&{}& \quad \quad \cdot \frac{\delta}{\delta T_{a_1b_1c_1}}\frac{\delta}{\delta \bar{T}_{d_1e_1f_1}} \frac{\delta}{\delta T_{a_2b_2c_2}}\frac{\delta}{\delta \bar{T}_{d_2e_2f_2}}. 
\eea
This yields
\bea
\Pi_{\mathcal{O}_{4,1}^{2,1}} \Gamma_{k\, 4,1}&=& { \bar g}_{4,1}^{2,1} + \frac{{ \bar g}_{4,1}^{2,2}}{ N^6}\left(N^5+ N^4\right)+ \frac{{ \bar g}_{4,1}^{2,3}}{ N^6}\left(N^5+ N^4\right), \nonumber\\
&{}& = { \bar g}_{4,1}^{2,1} + \mathcal{O}(1/N).
\eea
Similarly, by a simple exchange of the distinguished color, we can define $\Pi_{\mathcal{O}_{4,1}^{2,2}}$ and $\Pi_{\mathcal{O}_{4,1}^{2,3}}$, which provide a unique projection onto ${ \bar g}_{4,1}^{2,2}$ and ${ \bar g}_{4,1}^{2,3}$, respectively.

At higher order in the vertex expansion, the distinct contraction patterns of tensors in the different invariants again allow us to find projections which uniquely identify a given invariant at leading order in $1/N$.

Note that operators which are not invariants, and which are generated on the right-hand-side of the Wetterich-equation due to the breaking of the symmetry by the regulator, are not removed in our projection prescription. Thus, we project onto the coupling of interest plus a contamination from symmetry-breaking terms.

\section{Fixed-points in a $\left(T \bar{T}\right)^3$ truncation}
We set up a truncation order by order in the tensors. While the combinatorical structure has an impact on the canonical dimension, it is mainly determined by the number of tensors in an interaction: For every additional tensor or its conjugate in the interaction, we associate an extra factor of $ \sqrt{N}^{-1}$. Thus, starting from a dimensionless wave-function renormalization, the couplings of the $(T\bar{T})^2$ operators must have at least dimension -2. Every extra trace adds an extra factor of $N$, lowering the canonical dimension of the double-trace couplings by one in comparison to the single-trace couplings. Further, we observe that the combinatorial structure of the interactions can further lower the canonical dimensionality. 
As we will discover using the FRG, the melonic interactions seem to be those with the lowest canoncial dimensionality at any order in the tensors. 

\begin{figure}[!b]
\includegraphics[width=\linewidth]{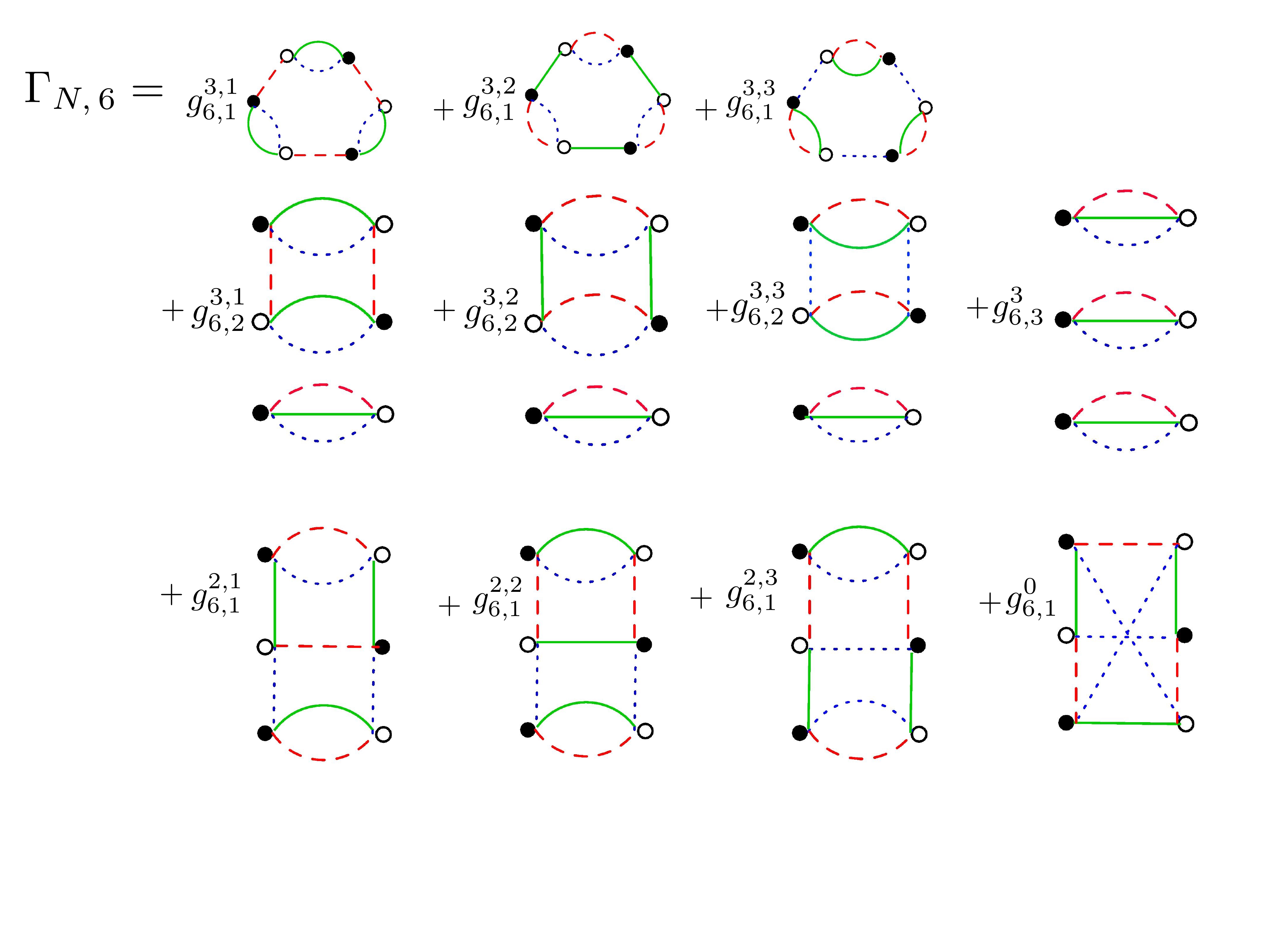}
\caption{\label{fig:T6trunc} The interactions in $\Gamma_6$ are sorted according to decreasing canonical dimensionality. All interactions apart from those in the last line are cyclic melons. The first three terms in the last line are melonic, the last one is not.}
\end{figure}

\begin{figure}[!t]
\includegraphics[width=0.7\linewidth]{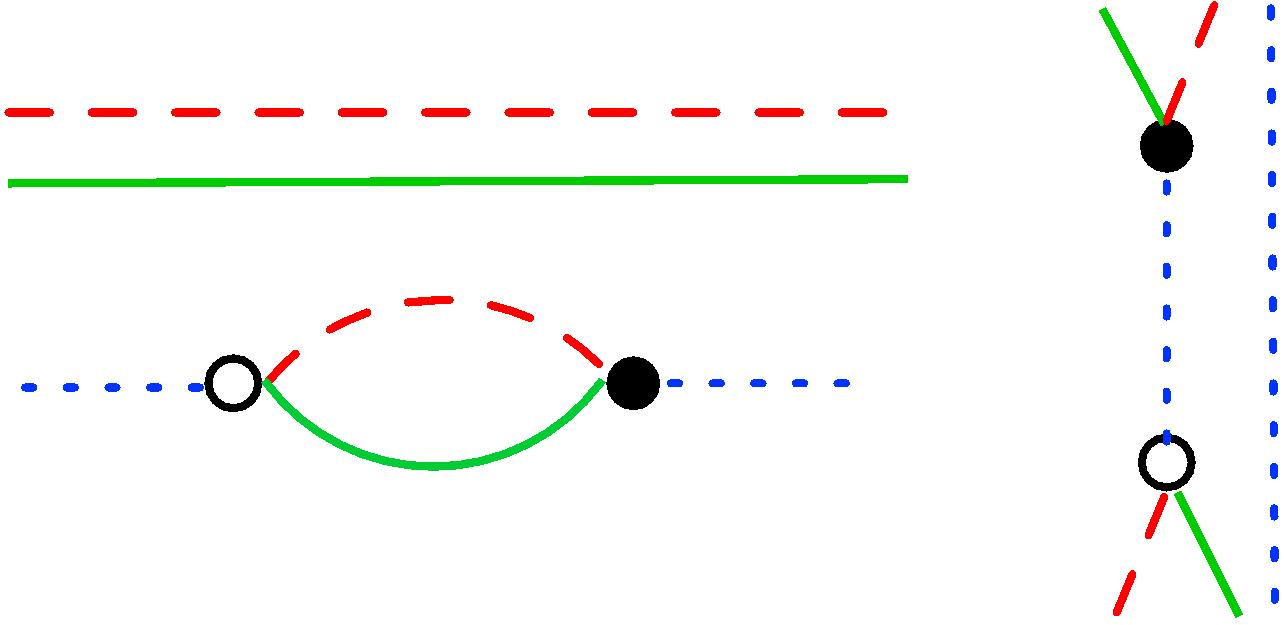}\\
\vspace{1.3cm}
\includegraphics[width=0.7\linewidth]{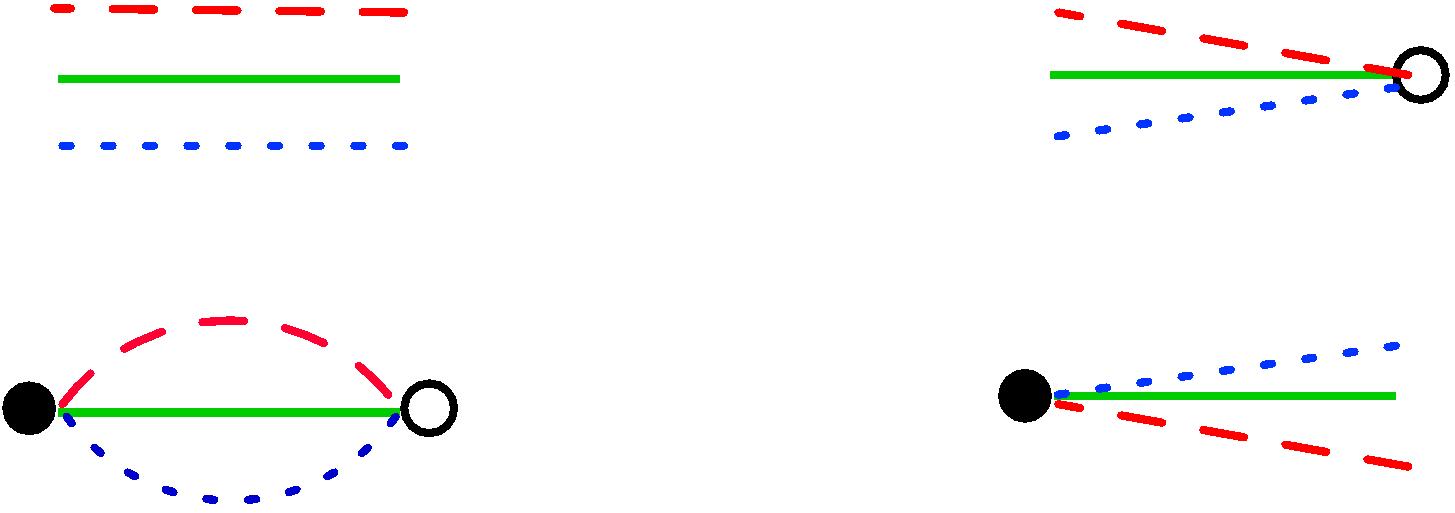}
\caption{\label{fig:Gamma24} $\Gamma^{(2)}_N\Big|_{g_{4,1}^{2,1}}$ (upper panel) contains two terms, one that still contains a melon, and serves as a building block for further melonic invariants, and one that contains ``two halves of slided melon" and that can be used to construct non-melonic invariants as well. $\Gamma^{(2)}_N\Big|_{g_{4,2}^{2}}$ (lower panel) contains two terms; one that contains a closed $T \bar{T}$-interaction and generates further disconnected interactions, and the second, which can be used to construct non-melonic invariants as well.}
\end{figure}

The first appearance of melonic, non-cylic invariants, as well as of non-melonic ones is at order $(T \bar{T})^3$. We will analyze a complete truncation at that order, which in addition to those invariants in \eqref{eq:simplesttrunc} and Fig.~\ref{fig:simplesttrunc} features, cf.~Fig.~\ref{fig:T6trunc}.
\bea
\Gamma_{N,\, 6}&=& \bar{g}_{6,1}^{3,1}\, T_{abc}\bar{T}_{ade}T_{fde}\bar{T}_{fgh}T_{igh}\bar{T}_{ibc}\nonumber\\
&{}& +\bar{g}_{6,1}^{3,2}\, \dots \quad \quad + \bar{g}_{6,1}^{3,3}\,\dots \nonumber\\
&{}& + \bar{g}_{6,1}^{2,1}\, T_{abc}\bar{T}_{dbe}T_{dfg}T_{hie}\bar{T}_{hig}\bar{T}_{afc}\nonumber\\
&{}& + \bar{g}_{6,1}^{2,2}\,\dots \quad \quad g_{6,1}^{2,3}\, \dots \nonumber\\
&{}& + \bar{g}_{6,1}^{0}\,T_{abc}\bar{T}_{ade}T_{fdg}\bar{T}_{hbg }T_{hie}\bar{T}_{fic}\nonumber\\
&{}&+\bar{g}_{6,2}^{3,1} \,T_{abc}\bar{T}_{ade}T_{fde}\bar{T}_{fbc}\, T_{ijk}\bar{T}_{ijk}\nonumber\\
&{}& +\bar{g}_{6,2}^{3,2}\,\dots \quad \quad + g_{6,2}^{3,3}\,\dots \nonumber\\
&{}& + \bar{g}_{6,3}^{3}\, T_{abc}\bar{T}_{abc}\, T_{def}\bar{T}_{def}\, T_{ghi}\bar{T}_{ghi},
\label{eq:T6trunc}
\eea
 where the ellipsis stand for the obvious change of preferred index.
To understand how these interactions are generated from the lower-order ones, consider that $\Gamma^{(2)}_{N}$ is obtained by removing a tensor $T$ and a complex conjugate $\bar{T}$ from the invariants, leaving the corresponding tensor indices open. The trace on the right-hand-side of the flow equation then glues these open legs together, respecting the colors. For instance, the entry in $\Gamma_N^{(2)}$ proportional to $g_{4,1}^{2,1}$ features one term with a closed melon, and one with ``two pieces of sliced melon", cf.~Fig.~\ref{fig:Gamma24}.
Thus, the combination $g_{4,1}^{2,1}\cdot g_{4,1}^{2,2}\cdot g_{4,2}^2$ contains one term which generates $g_{6,1}^{2,3}$.

We obtain the following beta-functions for the dimensionless counterparts of the couplings in our truncation
\bea
\eta&=& \frac{1}{20} \left(g_{4,1}^{2,1}+ g_{4,1}^{2,2}+ g_{4,1}^{2,3}+g_{4,2}^2\right) (5- \eta), \label{eq:fulltrunceta}\\
\beta_{g_{4,1}^{2,i}}&=& (2+2\eta) g_{4,1}^{2,i} + (g_{4,1}^{2,i})^2 \frac{13}{630}(21-4\eta) \nonumber\\
&{}& - g_{6,1}^{3,1} \frac{5-\eta}{15} -g_{6,2}^{3,1} \frac{5-\eta}{40}\label{eq:fulltruncbetag4}\\
\beta_{g_{4,2}^{2}}&=& (3+2 \eta)g_{4,2}^{2}, \label{eq:fulltruncbetag42}\\
&{}&+\frac{6- \eta}{15} \Bigl(\left(g_{4,2}^2\right)^2   +2 g_{4,2}^2\, \left(g_{4,1}^{2,1}+g_{4,1}^{2,2}+g_{4,1}^{2,3} \right) \nonumber\\
&{}& \quad \quad\quad\quad+ 2\,g_{4,1}^{2,1} \, g_{4,1}^{2,2}+ 2\,g_{4,1}^{2,1}\, g_{4,1}^{2,3}+2\, g_{4,1}^{2,2}\, g_{4,1}^{2,3}\Bigr) \nonumber\\
&{}&  - \frac{5-\eta}{20} \left(g_{6,2}^{3,1}+ g_{6,2}^{3,2}+ g_{6,2}^{3,3} \right).
\eea
The equation for the anomalous dimension can be solved to give
\be
\eta=\frac{5 \left(g_{4,1}^{2,1}+g_{4,1}^{2,2}+g_{4,1}^{2,3} +g_{4,2}^{2}\right)}{20 + g_{4,1}^{2,1}+g_{4,1}^{2,2}+g_{4,1}^{2,3}+g_{4,2}^2}.\label{eq:etasolsimplesttrunc}
\ee
 The beta functions for the higher-order couplings in our truncation are given by
\bea
\beta_{g_{6,1}^{3,1}}&=&(4+ 3 \eta)\, g_{6,1}^{3,1} +\frac{13}{210}(21-4 \eta) g_{4,1}^{2,1}\, g_{6,1}^{3,1}\nonumber\\
&{}&-8 \left(g_{4,1}^{2,1}\right)^3 \frac{5769 - 1049 \eta}{60480},
\eea
\bea
\beta_{g_{6,1}^{2,1}}&=&(5+3 \eta)\, g_{6,1}^{2,1}-  g_{4,1}^{2,1}\, g_{4,1}^{2,2}\, g_{4,1}^{2,3}\, 16 \frac{93869-15729 \eta}{362880}\nonumber\\
&{}& - \left(g_{4,1}^{2,2} \left(g_{4,1}^{2,3}\right)^2+g_{4,1}^{2,3} \left(g_{4,1}^{2,2}\right)^2\right) 8 \frac{46500-8887\eta}{151200}\nonumber\\
&{}& +\left(g_{4,1}^{2,2}g_{6,1}^{3,3}+ g_{4,1}^{2,3}g_{6,1}^{3,2}\right) 13\frac{21-4\eta}{210}\nonumber\\
&{}& + \left(g_{4,1}^{2,2}+g_{4,1}^{2,3} \right)g_{6,1}^{2,1} 13 \frac{21-4\eta}{630},
\eea
\bea
\beta_{g_{6,1}^{0}}&=& (-d_{g_{6,1}^0}+ 3 \eta)\, g_{6,1}^{0}\nonumber\\
&{}&- g_{4,1}^{2,1}\, g_{4,1}^{2,2}\, g_{4,1}^{2,3}\, 16 \frac{73160-13889 \eta}{604800}  \frac{N^{-d^{\bar{g}_{6,1}^0}}}{N^6},
\eea
\bea
\beta_{g_{6,2}^{3,1}}&=& (5+3 \eta)\, g_{6,2}^{3,1}+\left(g_{4,1}^{2,2}+{g_{4,1}^{2,3}}\right)g_{6,1}^{3,1}\frac{6-\eta}{5}\nonumber\\
&{}&-\left(g_{4,1}^{2,1}\right)^2\, \left(g_{4,1}^{2,2}+{ g_{4,1}^{2,3}}\right) 8\frac{2764-467 \eta}{10080}\nonumber\\
&{}& + \left(g_{4,1}^{2,2}+ g_{4,1}^{2,3}\right)g_{6,2}^{3,1}  \frac{6-\eta}{15}\nonumber\\
&{}&+g_{6,1}^{3,1}g_{4,2}^2 \frac{6- \eta}{5}  \nonumber\\
&{}&+ g_{6,2}^{3,1} g_{4,2}^2 \frac{6-\eta}{15}\nonumber\\
&{}& + g_{6,2}^{3,1}g_{4,1}^{2,1} \frac{399-73 \eta}{315},
\eea
\bea
\beta_{g_{6,3}^3}&=&(6+3 \eta)\, g_{6,3}^{3} +\Bigl(g_{4,1}^{2,2}\left[ g_{6,2}^{3,1}+g_{6,2}^{3,3}\right]+g_{4,1}^{2,1}\left[ g_{6,2}^{3,2}+g_{6,2}^{3,3}\right]\nonumber\\
&{}&+g_{4,1}^{2,3}\left[ g_{6,2}^{3,1}+g_{6,2}^{3,2}\right]\Bigr)\frac{6-\eta}{15}2\nonumber\\
&{}& +g_{6,3}^3 \left(g_{4,1}^{2,1}+g_{4,1}^{2,2}+g_{4,1}^{2,3} \right) \frac{6-\eta}{5}\nonumber\\
&{}& - g_{4,1}^{2,1}\, g_{4,1}^{2,2}\, g_{4,1}^{2,3} \frac{7-\eta}{84} 16\nonumber\\
&{}&+ g_{4,2}^2 g_{6,3}^3 \frac{6-\eta}{5} -8 \left(g_{4,2}^2\right)^3 \frac{7-\eta}{84}\nonumber\\
&{}& -\left(g_{4,2}^2\right)^2\left(g_{4,1}^{2,1}+g_{4,1}^{2,2}+g_{4,1}^{2,3} \right) 2\frac{7-\eta}{7}\nonumber\\
&{}& - \left(g_{4,1}^{2,1}\left[g_{4,1}^{2,2}+g_{4,1}^{2,3} \right] +g_{4,1}^{2,2}g_{4,1}^{2,3}\right)g_{4,2}^2 2 \frac{7-\eta}{7}.
\eea
Corresponding equations hold for the couplings with a preferred second or third index under appropriate permutations of all couplings.

Interestingly, it turns out that there is no contribution $\sim g_{6,1}^{0}$ to any of the beta functions at leading order in $1/N$  in our truncation. We find 
\be
d_{g_{6,1}^0} \geq -6. \label{eq:g610bound}
\ee
Due to the decoupling, the fixed-point search does not need to include $ g_{6,1}^{0}$, as it always automatically features a fixed point at 
\be
g_{6,1\,\,\ast}^{0}= \frac{73160-13889 \eta_{\ast}}{604800} \frac{16}{-d_{\bar{g}_{6,1}^0}+3 \eta_{\ast}} {g_{4,1\, \ast}^{2,1}\, g_{4,1\, \ast}^{2,2}\, g_{4,1\, \ast}^{2,3}}.
\ee
The corresponding critical exponent is given by
\be
\theta= d_{\bar{g}_{6,1}^0}- 3 \eta_{\ast}.\label{eq:thetag610}
\ee
As a consequence of the decoupling, our assignment of dimensionality is not unique: We could choose a less negative canonical dimensionality for that coupling than the one that saturates the bound \eqref{eq:g610bound} such that a nontrivial backcoupling into some of the beta functions would exist. On the other hand, the terms that are present within the beta function for $\beta_{g_{6,1}^0}$ in the present assignment would then vanish. Therefore, the coupling could be set to zero consistenly on all scales for that alternative choice of canonical dimension. As a consequence, one would obtain a fixed point at $g_{6,1\, \ast}^0=0$ with the same critical exponent as in \eqref{eq:thetag610}. 

Since the existence of a real fixed point for $g_{6,1}^0$ is guaranteed as soon as the other couplings assume real fixed-point values, and the corresponding critical exponent is clearly negative for all admissable values of $\eta$, we will neglect $g_{6,1}^0$ in our fixed point analysis.

The canonical dimensionality for all other couplings is fixed if we insist that the $1/N$ expansion is well-defined. In accordance with the result that the melonic interactions dominate the large-$N$ limit  \cite{Gurau:2010ba,Gurau:2011aq,Gurau:2011xq}, the melonic interactions have the largest canonical dimensionality at each order in $\left(T \bar{T} \right)^n$ and at fixed number of traces. In particular, the single-trace cyclic melons are the canonically leading operators at each order in $\left(T \bar{T} \right)^n$. Thus, the flow equation independently hints at the dominance of melons in the leading order in $1/N$, as this is the only assignment of dimensionality consistent with a well-defined $1/N$ expansion of the flow equation.

\subsection{Fixed-point search and critical exponents}
 We will discuss how to tentatively distinguish truncation-induced fixed points from actual ones. Moreover, we will explain different schemes for obtaining the critical exponents.

Eq.~\eqref{eq:etasolsimplesttrunc} for the anomalous dimension has a non-perturbative structure in the quartic couplings, as couplings appear in the denominator, leading to a Taylor expansion containing arbitrarily high powers in the couplings.  Thus the non-trivial denominator can induce additional zeros. These are generically nonperturbative, and thus a truncation scheme relying on the canonical dimensionality  might not reliably describe these fixed points. We will therefore discard  them and only focus on those fixed points which arise ``semi-perturbatively", i.e., from the polynomial structure of the beta-functions. We will therefore apply two simplifications to the beta-functions and only analyze fixed points which persist under both simplification steps: 

The first consists in a semi-perturbative approximation, in which all $\eta$'s that arise from the scale-derivative of the regulator are neglected.  This leaves those factors of $\eta$ that come paired with the canonical dimensionality, and arise from the definition of the dimensionless coupling in \eqref{eq:dimlesscoup}.
The same logic has been applied, e.g.,  in the context of asymptotically safe gravity in \cite{Dona:2015tnf}.
This yields the following expression for the anomalous dimension
\be
\eta_{\rm semi-pert}= \frac{5}{20}\left(g_{4,1}^{2,1}+g_{4,1}^{2,2}+g_{4,1}^{2,3}+g_{4,2}^2 \right).
\ee

The second approximation, which we call the perturbative one consists in setting $Z= 1$ and therefore $\eta=0$.  Checking whether a fixed point persists with limited quantitative changes as we go from the full result to the first and second approximation can be understood as checking the stability of the fixed point under changes of the truncation.

We define the critical exponents as minus the eigenvalues of the stability matrix  of the couplings in our truncation, i.e., 
\be
\theta_I = -{\rm eig}\, \mathcal{M}_{ij}= - {\rm eig} \left(\frac{\partial \beta_{g_i}}{\partial g_j} \right)\Big|_{\vec{g}= \vec{g}_{\ast}},
\ee
such that they agree with the canonical dimensionality for the non-interacting fixed point.  Note that, the model cannot become asymptotically free, as all couplings have negative dimensionality, i.e., they correspond to UV repulsive directions of the free fixed point.
Herein, we have summarized the different couplings in one vector which is labelled by one index, i.e., 
\bea
\vec{g}^T &=&(g_{4,1}^{2,1}, g_{4,1}^{2,2}, g_{4,1}^{2,3}, g_{4,2}^2, g_{6,1}^{3,1}, g_{6,1}^{3,2}, g_{6,1}^{3,3}, g_{6,1}^{2,1}, g_{6,1}^{2,2}, g_{6,1}^{2,3},\nonumber\\
&{}& g_{6,2}^{3,1}, g_{6,2}^{3,2}, g_{6,2}^{3,3}, g_{6,3}^3). 
\eea

For the critical exponents, we will compare two different schemes: In the first scheme, the derivative of the anomalous dimension with respect to the couplings is taken into account in the stability matrix, i.e., 
\be
\mathcal{M}_{ij}=\left( \frac{\partial \beta_{g_i}}{\partial g_{j}}+ \sum_k \frac{\partial \beta_{g_i}}{\partial \eta_k} \frac{\partial \eta_k}{\partial g_j}\right)\Big|_{\vec{g} =\vec{g}_{\ast}}.\label{eq:theta}
\ee
In the second prescription, we hold $\eta=\rm const$ to evaluate the entries of the stability matrix:
\be
\mathcal{M}_{ij}'=\left( \frac{\partial \left(\beta_{g_i}|_{\eta= \eta_{\ast}}\right)}{\partial g_{j}}\right)\Big|_{\vec{g} =\vec{g}_{\ast}}.\label{eq:thetaprime}
\ee
We will denote the critical exponents obtained from the second prescription by $\theta'$.

The second prescription has been discussed in detail in \cite{Boettcher:2015pja} for the application of the FRG in the context of multicritical phenomena for models with $O(N)\oplus O(M)$ symmetry, see \cite{Eichhorn:2013zza}: There, interacting fixed points emanate from the Gau\ss{}ian fixed point to become interacting below $d=4$, and can thus be followed to $d=3$ employing the $\epsilon$ expansion. For one of the critical exponents of a particular, decoupled fixed point, a scaling relation holds order by order in the $\epsilon$ expansion \cite{Aharony}.  It then turns out that the FRG in a local potential approximation with anomalous dimensions only respects the scaling relation when the second prescription for the critical exponents is employed. This is presumably linked to the fact that the anomalous dimension only arises at two-loop order in the perturbative expansion, while the beta functions obtained from the FRG in that truncation are only one-loop exact, not two-loop exact. Therefore the first prescription for the critical exponents actually leads to a larger deviation form the exact result than the second. Of course, the first prescription would be  exact in an untruncated theory space.

\subsection{Fixed point with one relevant direction: Double-scaling limit from the FRG}
The double scaling limit requires taking $N \rightarrow \infty$ while tuning one of the couplings. According to \cite{Dartois:2013sra, Bonzom:2014oua} it features one relevant direction with $\theta= D-2$, where $D$ is the rank, i.e., $D=3$ in our case.. We discover a fixed point with corresponding qualitative features -- in particular just one relevant direction. This requires that the fixed-point values of all multi-trace couplings vanish, as all fixed points with nonvanishing multi-trace contributions feature additional relevant directions.  
This in turn is only possible if the fixed point does not exhibit a color symmetry; in fact, only melonic couplings with one preferred color can be non-zero. This can be seen directly by considering $\beta_{g_{4,2}^2}$, cf.~Eq.~\eqref{eq:fulltruncbetag42}: As soon as two of the single-trace quartic couplings have non-vanishing fixed-point values, $g_{4,2}^2=0$ no longer solves its fixed-point equation.

The fixed-point properties are shown in tab.~\ref{tab:FPsingletrace} and tab.~\ref{tab:thetasingletrace}. 

\begin{table}[!b]
\begin{tabular}{c||c|c|c|c|c|c|}
 scheme & $g_{4,1\, \ast}^{2,1}$&$g_{4,2\, \ast}^{2}$&$g_{6,1\, \ast}^{3,1}$&$g_{6,1\, \ast}^{2,1}$&$g_{6,2\, \ast}^{3,1}$&$g_{6,3\, \ast}^3$\\ \hline \hline
full & -1.94& 0 & --& -- &--& -- \\ \hline
semi-pert & -2.14& 0 &--& -- &--& -- \\ \hline
pert & -4.62& 0 &--& -- &--& -- \\ \hline \hline
full & -1.37 & 0 & -2.14 & 0 & 0 & 0 \\ \hline
semi-pert & -1.47 & 0 & -2.46 & 0 & 0 & 0 \\\hline
pert &-2.14 & 0 & -6.12 & 0 &0 & 0 \\\hline \hline
\end{tabular}
\caption{\label{tab:FPsingletrace} Fixed-point values of the couplings. All couplings which are not shown explicitly are understood to be zero. Couplings which are not included in a truncation are indicated by a dash. Within each truncation, the properties of the fixed point are given in the case with the full anomalous dimensions, in the semi-perturbative approximation, in which all $\eta$s that arise when the scale-derivative acts on the regulator are set to zero, and in the perturbative approximation, where no anomalous dimension is taken into account.}
\end{table}

 At first sight, a fixed point that is not color symmetric does not seem to be connected with the color symmetric double scaling limit. However, we point out that this may be an artifact of our truncation: The emergence of fixed points that distinguish one particular index position might be due to the fact that we use a finite Taylor expansion in  "Euclidean" coordinates on theory space, which breaks the rotational symmetry that would be manifest in "spherical" coordinates. In particular a finite Taylor expansion of $r=\sqrt{x_1^2+x_2^2+x_3^2}$ can break rotational symmetry into a discrete symmetry of permutations among the coordinates $x_i$. In this respect, we note that the global exchange symmetry between the three distinct indices (i.e., the colors in the graphical representation of the interactions) is satisfied as three versions of this fixed point exists, which are obtained by the exchange of the three colors. This might hint that the true fixed point that we find is indeed color symmetric. In that case, the fixed point should be visible in a different choice of basis in theory space, spanned by a color-symmetric version of each coupling in addition to couplings that parameterize the possible deviations from color symmetry for the different combinatorial structures.

\begin{widetext}
\begin{table*}
\begin{tabular}{c||c|c|c|c|c|c|c|c|c|c|c|c|c|c|c|c|c||c|}
 scheme &$\theta_1$&$\theta_2$&$\theta_3$&$\theta_4$&$\theta_5$&$\theta_6$&$\theta_7$&$\theta_8$&$\theta_9$&$\theta_{10}$&$\theta_{11}$&$\theta_{12}$&$\theta_{13}$&$\theta_{14}$&$\eta$\\ \hline \hline
 full & 2.21 & -0.24&-0.93 & -0.93&--&--&--&--&--&--&--&--&--&--&-0.54\\ \hline
 semi-pert & 2 & -0.21 & -0.68 & -0.68&--&--&--&--&--&--&--&--&--&--&-0.54\\ \hline
 pert & 2 & 0.69 & -2&-2&--&--&--&--&--&--&--&--&--&--&-0.54\\ \hline \hline
  full &2.14 & -0.59 & -1.26 & -1.26 &-2.24 &-2.54&-2.89 & -2.89 &-3.15&-3.26 &-3.26 & -3.31& -3.31 & -3.89 &-0.37 \\ \hline 
  semi-pert & 2 & -0.58 & -1.26 & -1.26 & -2.24 & -2.54 & -2.90 & -2.90 &-3.13&-3.26 & -3.26 & -3.31& -3.31&-3.90 &-0.37 \\ \hline
  pert & 2 & -0.35 & -2&-2&-3.23 & -3.37 &-3.43& -4&-4 & -4.07 & -4.07 & -4.14& -4.14& -5  & 0 \\\hline  \hline
\end{tabular}\newline\\
\vspace{0.5cm}
\begin{tabular}{c||c|c|c|c|c|c|c|c|c|c|c|c|c|c|c|c|c||c|}
 scheme &$\theta_1'$&$\theta_2'$&$\theta_3'$&$\theta_4'$&$\theta_5'$&$\theta_6'$&$\theta_7'$&$\theta_8'$&$\theta_9'$&$\theta_{10}'$&$\theta_{11}'$&$\theta_{12}'$&$\theta_{13}'$&$\theta_{14}'$&$\eta$\\ \hline \hline
 full & 0.93 & -0.24&-0.93 & -0.93&--&--&--&--&--&--&--&--&--&--&-0.54\\ \hline
 semi-pert & 0.93 & -0.21 & -0.93 & -0.93&--&--&--&--&--&--&--&--&--&--&-0.54\\ \hline
 \hline
  full &1.33 & -0.59 & -1.26&-1.26  &-2.30 &-2.54&-2.89 & -2.89 &-3.15&-3.26 &-3.26 & -3.31& -3.31 & -3.89 &-0.37 \\ \hline 
  semi-pert & 1.33 & -0.58 & -1.26 & -1.26 & -2.31 & -2.54 & -2.90 & -2.90 &-3.13&-3.26 & -3.26 & -3.31& -3.31&-3.90 &-0.37 \\ \hline
  \hline
\end{tabular}\newline\\
\caption{\label{tab:thetasingletrace} Critical exponents at the fixed point in tab.~\ref{tab:FPsingletrace} in the prescription according to Eq.~\eqref{eq:theta} (upper table) and according to Eq.~\eqref{eq:thetaprime} (lower table).  Within each truncation, the properties of the fixed point are given in the case with the full anomalous dimensions, in the semi-perturbative approximation, in which all $\eta$s that arise when the scale-derivative acts on the regulator are set to zero, and in the perturbative approximation, where no anomalous dimension is taken into account.}
\end{table*}
\end{widetext}

The multiplicity of some of the irrelevant critical exponents arises from the discrete symmetry-enhancement in the theory space: There is an exchange symmetry between those couplings  belonging to the same combinatorical structure of tensor contractions, which are set to zero at the fixed point, e.g., $ g_{4,1}^{2,2},  g_{4,1}^{2,3}$. There are two eigendirections of the stability matrix associated to those couplings, which are on an equal footing, i.e., the critical exponents associated to those directions are equal. The same pattern must persist at higher orders in the truncation, whenever there are three different couplings associated to the same combinatorial structure of the tensors, only one of which is nonzero. Note that although the couplings $g_{4,1}^{2,2},  g_{4,1}^{2,3}$, and so on are set to zero at the fixed point, the interactions in the third sector, associated to $g_{4,1}^{2,1}$ are sufficient to generate nontrivial critical exponents, just as discussed in detail in \cite{Eichhorn:2013zza}.

Under an enlargement of the truncation from quartic to hexic, the fixed-point value for the quartic coupling changes by about $30 \%$. Encouragingly, the difference between the fixed-point estimates for the leading-order coupling in the different schemes decreases under the enlargement of the truncation, as it should be expected for an actual fixed point. 

The negative fixed-point values might be interpreted as pointing towards a phase of broken symmetry. Specifically, the model is invariant under phase rotations of the tensors, where
\bea
T_{ijk} \rightarrow e^{i\, \alpha}T_{ijk},\\
\bar{T}_{ijk}\rightarrow e^{-i \alpha}\bar{T}_{ijk}.
\eea
The negative fixed-point values of the couplings hint towards either an unstable potential, or towards a non-trivial minimum that would break this symmetry. The first case cannot be identified reliably when a polynomial expansion of the potential is used. The second case would suggest that the fixed-point values and critical exponents can be determined more reliably by expanding around the nontrivial mininum of the potential. We will defer the study of the corresponding parameterization of the potential to future work, and merely remark, that this could be a reason why our critical exponents are quantitatively imprecise.

We observe a significant difference between the leading critical exponent in the two schemes. The second scheme, in which the critical exponents do not include contributions from the derivative of the anomalous dimension, gives a result significantly closer to $\theta_1=1$, as we would expect for the double-scaling limit according to \cite{Dartois:2013sra,Bonzom:2014oua}. The difference between the schemes is significantly less pronounced in the larger truncation. On the other hand, it is clear that these values have not yet converged to numerically reliable results. We  tentatively conclude that our results  are 
consistent with the interpretation that we redisover the scaling underlying the double-scaling limit with our method.

\subsection{Fixed point with two relevant directions: Beyond double-scaling}
\begin{table}[!t]
\begin{tabular}{c||c|c|c|c|c|c|}
 scheme & $g_{4,1}^{2,1}$&$g_{4,2}^{2}$&$g_{6,1}^{3,1}$&$g_{6,1}^{2,1}$&$g_{6,2}^{3,1}$&$g_{6,3}^3$\\ \hline \hline
full & -1.05& -1.33 & --& -- &--& -- \\ \hline
semi-pert& -1.58 & -1.05& --& -- &--& -- \\ \hline
pert & -4.62 & 1.73 & --& -- &--& -- \\ \hline\hline
full & -0.53 & -2.11 & -0.14 &0&-0.39 &-- \\ \hline
semi-pert & -0.63 & -2.35 & -0.20 & 0&-0.57 & -- \\\hline
pert & -2.01 & -1.64 & -4.43 &0 & -4.84 & -- \\\hline\hline
full & -1.04 & -0.84 &-0.97 & 0 & -0.64 & -0.99 \\ \hline
semi-pert & -1.15 & -0.89 & -1.21& 0 &-0.78 & -1.14 \\ \hline
pert & -2.10&-0.54 &-5.54 &0 &-1.68 &-0.47 \\ \hline\hline
\end{tabular}
\caption{\label{tab:FPmultitrace} Fixed-point values of couplings and critical exponents. All couplings which are not shown explicitly are understood to be zero.}
\end{table}

\begin{widetext}
\begin{table*}[!t]
\begin{tabular}{c||c|c|c|c|c|c|c|c|c|c|c|c|c|c|c|c|c||c|}
 scheme &$\theta_1$&$\theta_2$&$\theta_3$&$\theta_4$&$\theta_5$&$\theta_6$&$\theta_7$&$\theta_8$&$\theta_9$&$\theta_{10}$&$\theta_{11}$&$\theta_{12}$&$\theta_{13}$&$\theta_{14}$&$\eta$\\ \hline \hline
 full & 2.64 & 0.14&-0.65 & -0.65&--&--&--&--&--&--&--&--&--&--&-0.67\\ \hline
 semi-pert & 2.28 &0.14 & -0.68 & -0.68& --&--&--&--&--&--&--&--&--&--&-0.66\\ \hline
 pert & 2& -0.69 & -2 & -2&--&--&--&--&--&--&--&--&--&--&0\\ \hline \hline
 full & 3.10&0.25 &-0.48 & -0.48 &  \multicolumn{2}{c|}{-1.22 $\pm$ i 0.44}&-1.53 & -1.53&-1.72& -1.72&-2.46 & -2.46 & -2.72&--&-0.76 \\ \hline 
 semi-pert &2.68 & 0.26 & -0.51 & -0.51 & \multicolumn{2}{c|}{-1.24 $\pm$ i 0.45}&-1.57 & -1.57 &-1.76 & -1.76 &-2.49 & -2.49 & -2.76 & -- & -0.75 \\ \hline
 pert & 2.11 & 0.34& -2&-2& -2.99 & -2.99 & -3.54 & -3.54 & -4 & -4 &-4.13&-4.13 &-5&--&0 \\ \hline  \hline
full& 2.56 & 0.44 &-0.97 & -0.97 &\multicolumn{2}{c|}{-1.80 $\pm$ i 0.42}& -2.45 & -2.45 & -2.63 & -2.63&-2.95 & -2.95& -3.06 & -3.45&-0.52 \\ \hline
semi-pert & 2.30 & 0.44 & -0.98 & -0.98 & \multicolumn{2}{c|}{-1.81 $\pm$ i 0.42}& -2.47 & -2.47 & -2.65 & -2.65 & -2.97 & -2.97&-3.06 & -3.47 &-0.51 \\ \hline
pert & 2.04 &0.37 & -2 & -2& -2.68 & -3.16 & -3.87 & -3.95 & -3.95 &-4 & -4 & -4.09 & -4.09& -5 & 0 \\ \hline
 \hline 
\end{tabular}\newline\\
\vspace{0.5cm}
\begin{tabular}{c||c|c|c|c|c|c|c|c|c|c|c|c|c|c|c|c|c||c|}
 scheme &$\theta_1'$&$\theta_2'$&$\theta_3'$&$\theta_4'$&$\theta_5'$&$\theta_6'$&$\theta_7'$&$\theta_8'$&$\theta_9'$&$\theta_{10}'$&$\theta_{11}'$&$\theta_{12}'$&$\theta_{13}'$&$\theta_{14}'$&$\eta$\\ \hline \hline
 full & 0.65 & 0.47&-0.65 & -0.65&--&--&--&--&--&--&--&--&--&--&-0.67\\ \hline
 semi-pert & 0.68 &0.42 & -0.68 & -0.68& --&--&--&--&--&--&--&--&--&--&-0.66\\ \hline \hline
 full & 1.03&0.43 &-0.48 & -0.48 &  \multicolumn{2}{c|}{-1.23 $\pm$ i 0.46}&-1.53 & -1.53&-1.72& -1.72&-2.46 & -2.46 & -2.72&--&-0.76 \\ \hline 
 semi-pert &1.04 & 0.44 & -0.51 & -0.51 & \multicolumn{2}{c|}{-1.25 $\pm$ i 0.47}&-1.57 & -1.57 &-1.76 & -1.76 &-2.49 & -2.49 & -2.76 & -- & -0.75 \\ \hline  \hline
full& 1.17 & 0.64 &-0.97 & -0.97 &\multicolumn{2}{c|}{-1.82 $\pm$ i 0.42}& -2.45 & -2.45 & -2.63 & -2.63&-2.95 & -2.95& -3.10 & -3.45&-0.52 \\ \hline
semi-pert & 1.17 & 0.63 & -0.98 & -0.98 & \multicolumn{2}{c|}{-1.83 $\pm$ i 0.42}& -2.47 & -2.47 & -2.65 & -2.65 & -2.97 & -2.97&-3.10 & -3.47 &-0.51 \\ \hline
 \hline 
\end{tabular}\newline\\
\caption{\label{tab:thetamultitrace} Fixed-point values of the critical exponents and the anomalous dimension at the fixed point in Tab.~\ref{tab:FPmultitrace}. The critical exponents are evaluated according to Eq.~\eqref{eq:theta} (upper table) and according to Eq.~\eqref{eq:thetaprime} (lower table). In smaller truncations only a subset of the critical exponents are evaluated, the others are denoted by a dash.}
\end{table*}
\end{widetext}

Allowing multi-trace operators to feature non-zero fixed-point values induces a fixed point with a second relevant direction, cf.~tab.~\ref{tab:FPmultitrace} and tab.~\ref{tab:thetamultitrace}. We conjecture that this could be a way of taking the continuum limit beyond the double-scaling limit. Due to the presence of nonvanishing multi-trace operators, which correspond to disconnected chunks of spacetime, the geometric interpretation of this fixed point is less straightforward than that of the one with one relevant direction. The existence of degrees of freedom which are not part of the continuous geometry, but which contain disconnected bits might point towards a topologically nontrivial phase, 
or to the presence of further degrees of freedom in the continuum limit which could potentially be interpreted as matter degrees of freedom. 
In the simplest case this could be an additional scalar field with a $\mathbb{Z}_2$ symmetry, in order to exclude the simplest form of instability from the microscopic potential. For $\mathbb{Z}_2$ symmetric scalar field theory in 3 dimensions, a well-known interacting fixed point exists, the Wilson-Fisher fixed point, which features one relevant direction. Potentially, that fixed point survives under the coupling to quantum gravity,  see, e.g., \cite{Percacci:2015wwa,Borchardt:2015rxa},  in the simplest case adding one additional relevant direction to the spectrum of critical exponents. Whether the fixed point that we discover can indeed be interpreted in this manner remains an exciting open question at this stage.

On the other hand, it is interesting to compare the results for the critical exponents to those obtained at a UV fixed point in truncations of the RG flow for continuum quantum gravity based on metric variables, for reviews of the asymptotic safety scenario for that case see \cite{ASreviews}. The results on both sides, in particular for the tensor model, are insufficiently converged to make a quantitatively precise comparison. Here, we simply observe that the results in continuum quantum gravity within the Einstein-Hilbert truncation are $\theta_1=2.47$, $\theta_2=0.77$ \cite{Biemans:2016rvp}, which is not incompatible with our results for the critical exponents in tensor models, cf.~Tab.~\ref{tab:thetamultitrace}. Note however, that results in \cite{Rechenberger:2012pm} suggest that even in $d=3$, $R^2$ adds another relevant direction beyond the two from the Einstein-Hilbert action. If those results persist to higher order, then the fixed point that we have explored for the tensor model does not directly correspond to the same universality class, discovered in a different formulation of quantum gravity, as it lacks one relevant direction.

We can also compare our leading relevant exponent to the critical exponent for the Newton coupling, obtained in numerical studies of Regge gravity \cite{Hamber:1992df}, which is given by $\theta = 1/\nu\approx 1.7$. Again, our results are not quantitatively precise, but the value from lattice gravity lies right within the range $\theta_1 \in (1.17,2.56)$ that we obtain from the two different schemes for $\theta$ from our largest truncation.

It is encouraging to see that the difference between the leading critical exponents in the two schemes, $\theta_1 - \theta_1'$ and $\theta_2-\theta_2'$, decreases, as we enlarge the truncation. Moreover, the fixed-point values for the quartic couplings show a smaller difference between the full, semi-perturbative and perturbative scheme in the largest truncation, compared to the quartic one. Together with the results on the critical exponents this could be interpreted as a sign of stability of the fixed point, i.e., it does not show the characteristics expected of a truncation artifact, and might therefore exist in the full theory space.\newline\\

\subsection{Universality in the continuum limit}

\begin{table}[!t]
\begin{tabular}{c||c||c|c|c|c|c|c|}
global sym. & scheme & $g_{4,1}^{2,1}$&$g_{4,2}^{2}$&$g_{6,1}^{3,1}$&$g_{6,1}^{2,1}$&$g_{6,2}^{3,1}$&$g_{6,3}^3$\\ \hline \hline
yes & full & 0 &-2.88 & -- & -- &--& -- \\ \hline 
yes & semi-pert &0& -3.33 &  -- & -- &--& --\\
\hline
yes  & pert & 0 & -7.5 & -- & -- &--& --\\ \hline
\hline
yes  & full &0.36 & -4.38 & 0.03 & 0.20 & 0.43 &-- \\ \hline
 yes  & semi-pert & 0.44 & -5.25 & 0.04 & 0.68 & 0.29 &-- \\ \hline
yes  & pert & 1.38 & -18.01 & 0.34 &3.63& 27.59 & --\\ \hline\hline
yes  & full & 0.99 & -6.06 &0.31 &3.61 &2.84 & -16.41\\ \hline
yes  & semi-pert & 1.18 & -7.17 &0.45 & 4.10 &5.28 & -23.73 \\ \hline
yes  & pert & 6.54 & -27.44 & 17.09 & 210.51 & 201.79 &-694.99 \\ \hline \hline
none & full & 0 & -2.88 & -- & -- &--& -- \\ \hline 
none & semi-pert & 0 & -3.33 &-- & -- &--& -- \\ \hline
none & pert & 0&-7.5&-- & -- &--& -- \\ \hline
\hline
none & full&0.36 & -4.38 & 0.03 & 0.40 & 0.43 & -- \\ \hline
none& semi-pert & 0.44 & -5.25 & 0.04 & 0.68 & 0.29 &-- \\ \hline
none & pert & 1.38 & -18.01 &0.35 &3.63&27.59 &--\\\hline\hline
none & full & 0.99 & -6.06 &0.31 & 2.84&3.61 &-16.41 \\ \hline
none & semi-pert &1.18 & -7.17 & 0.45 & 4.10 & 5.28 & -23.73 \\ \hline
none & pert &6.54 & -27.44&17.09&210.51 &201.79 & -694.99 \\ \hline \hline
\end{tabular}
\caption{\label{tab:fpcolorfree}We show the fixed-point values in the theory space with global color symmetry and that without. For the latter case, we do not explicitly write the fixed-point values for all additional couplings, as they are given by the fully color-symmetric choice.}
\end{table}

\begin{table*}
\begin{tabular}{c||c||c|c|c|c|c|c|c|c|c|c|c|c|c|c|c|c|c||c|}
global symm. & scheme &$\theta_1$&$\theta_2$&$\theta_3$&$\theta_4$&$\theta_5$&$\theta_6$&$\theta_7$&$\theta_8$&$\theta_9$&$\theta_{10}$&$\theta_{11}$&$\theta_{12}$&$\theta_{13}$&$\theta_{14}$&$\eta$\\ \hline \hline
yes  & full & 3.47 & -0.32&-- & -- &-- & -- &-- & -- &-- & -- &-- & --&-- & --&-0.84 \\\hline 
yes & semi-pert & 3 & -0.33&-- & -- &-- & -- &-- & -- &-- & -- &-- & --&-- & --&-0.83 \\\hline
yes  & pert & 3 & -2&-- & -- &-- & -- &-- & -- &-- & -- &-- & --&-- & --&-0.83 \\\hline  \hline
yes  & full & 4.20 & 0.50 & \multicolumn{2}{c|}{-1.47 $\pm$ i 0.27}& -2.41 & -- & --&--&--&--&-- & --&-- & --&-0.99 \\ \hline
yes  & semi-pert & 3.63 & 0.52 &\multicolumn{2}{c|}{-1.49 $\pm$ i 0.27}& -2.43 &  -- & --&--&--&-- &--&--&-- & --&-0.98 \\ \hline
yes  & pert & 6.08  &1.85  &-3.71  & -5.76 &-6.19 &  -- & --&--&--&-- &--&--&-- & --&0 \\ \hline\hline
yes  & full &6.17 & 1.41 &-2.41 & \multicolumn{2}{c|}{-2.88 $\pm$ i 0.77} & -3.27 &--&--&-- &--&--&--&--&--&-0.91 \\ \hline
yes & semi-pert &5.69 & 1.32 &-2.49& \multicolumn{2}{c|}{-2.91 $\pm$ i 0.80}  &-3.31 & --&--&--&-- &--&--&--&--&-0.91 \\ \hline
yes  & pert & 15.32 & 1.94 & -10.67 & \multicolumn{2}{c|}{-11.80 $\pm$ i 8.54} & -14.76&--&--&--&-- &--&--&--&--&0
\\\hline\hline
none & full & 3.47 & -0.32 & -0.32 & -0.32&-- & -- &-- & -- &-- & -- &-- & --&-- & --&-0.84 \\ \hline 
none & semi-pert &3 & -0.33 & -0.33 & -0.33&-- & -- &-- & -- &-- & -- &-- & --&-- & --&-0.83 \\ \hline 
none & pert &2 & -2 & -2 & -2&-- & -- &-- & -- &-- & -- &-- & --&-- & --&0 \\ \hline 
\hline
none & full&4.20 & 0.50 & -0.025 &-0.025& \multicolumn{2}{c|}{-1.42 $\pm$ i 0.24} & \multicolumn{2}{c|}{-1.42 $\pm$ i 0.24} &\multicolumn{2}{c|}{-1.47 $\pm$ i 0.27}& -2.41  &-2.41 &-2.41& --& -0.99 \\ \hline
none& semi-pert & 3.63 & 0.52 &-0.02&-0.02&\multicolumn{2}{c|}{-1.44 $\pm$ i 0.24}&\multicolumn{2}{c|}{-1.44 $\pm$ i 0.24}&\multicolumn{2}{c|}{-1.49 $\pm$ i 0.27}& -2.43 &-2.43&-2.43&--&-0.98 \\ \hline
none & pert & 6.08 & 1.85 & -0.51 & -0.51 &-3.44&-3.44&-3.71&-5.68 & -5.68&-5.76 & -6.19 & -6.19 &-6.19 &--&--\\\hline\hline
none & full & 6.17 & 1.41 & -0.09 & -0.09 &-2.41 & \multicolumn{2}{c|}{-2.85 $\pm$ i 0.58}&\multicolumn{2}{c|}{-2.85 $\pm$ i 0.58}& \multicolumn{2}{c|}{-2.88 $\pm$ i 0.77}& -3.27 & -3.27 & -3.27 & - 0.91\\ \hline
none & semi-pert & 5.69 & 1.32 & -0.09 & -0.09 &-2.49& \multicolumn{2}{c|}{-2.90 $\pm$ i 0.58}& \multicolumn{2}{c|}{-2.90 $\pm$ i 0.58}& \multicolumn{2}{c|}{-2.91 $\pm$ i 0.80}&  -3.31 & -3.31 &-3.31 & -0.91 \\ \hline
none & pert & 15.32 & 1.94 & -0.81 & -0.81 & -10.67 & -10.67 & -10.67& \multicolumn{2}{c|}{-11.80 $\pm$ i 8.54}
& \multicolumn{2}{c|}{-13.46 $\pm$ i 1.00}& \multicolumn{2}{c|}{-13.46 $\pm$ i 1.00}&-14.76 & -- \\ \hline \hline
\end{tabular}\newline\\
\vspace{0.5cm}
\begin{tabular}{c||c||c|c|c|c|c|c||c|}
global sym. & scheme &$\theta_1'$&$\theta_2'$&$\theta_3'$&$\theta_4'$&$\theta_5'$&$\theta_6'$&$\eta$\\ \hline \hline
yes & full & 1.32 & -0.32&-- & -- &-- & -- &-0.84 \\\hline 
yes  & semi-pert & 1.33 & -0.33&-- & -- &-- & -- &-0.83 \\\hline
  \hline
yes  & full & 1.91 & 0.17 & \multicolumn{2}{c|}{-1.45 $\pm$ i 0.23}& -2.41 &--& -0.99 \\ \hline
 yes  & semi-pert & 1.95 & 0.19 &\multicolumn{2}{c|}{-1.47 $\pm$ i 0.23}& -2.43 &--& -0.98 \\ \hline
\hline
yes  & full &4.93 & 10.38 & \multicolumn{2}{c|}{-2.73 $\pm$ i 1.36}&-2.98 & -3.27 &-0.91 \\ \hline
yes  & semi-pert &5.04 & 0.39 & \multicolumn{2}{c|}{-2.76 $\pm$ i 1.41} &-3.03 &-3.31 &-0.91 \\ \hline
\hline
\end{tabular}\newline\\
\caption{\label{tab:thetacolorfree}We show the critical exponents in the theory space with global color symmetry and that without. For the critical exponents in the second prescription, $\theta_1'$, we restrict ourselves to the color-symmetric case, as we are only interested in the estimate of the positive critical exponents.}
\end{table*}

The distinction of the indices and the corresponding $U(N)\otimes U(N)\otimes U(N)$ symmetry, while necessary for a well-defined $1/N$ expansion to exist, does not have any obvious physical interpretation in the continuum limit.  In other words, it should be possible to find a fixed point such that the coloring of the edges in the graphs corresponding to the different interactions does not impact the continuum limit.
Thus, one could expect that the distinction of the different indices and the corresponding couplings (e.g., $g_{4,1}^{2,1}$ and $g_{4,1}^{2,2}$) should not matter for the discovery of a fixed point. In fact, we can confirm this hypothesis, by comparing critical exponents of fixed points in the model where the distinction of indices in the invariants leads to the distinction of couplings, and the model with a global exchange
symmetry, which maps the three colors onto each other. In that model, the couplings $g_{n,m}^{k,i}$ for the different $i$'s should be identified with each other, e.g., $g_{4,1}^{2,1}=g_{4,1}^{2,2}= g_{4,1}^{2,3}$. We then find a fixed point with two relevant directions when we take into account single- and double-trace terms up to $T^6$, cf.~Tab.~\ref{tab:fpcolorfree} and \ref{tab:thetacolorfree}. To understand whether the global symmetry affects that fixed point, and in particular the number of relevant directions, we search for the same fixed point in the extended theory space, 
where we distinguish the couplings. Setting the different couplings to the previous fixed-point values does of course lead to a fixed point with a degeneracy in the fixed-point values, cf.~Tab.~\ref{tab:fpcolorfree}. The existence of the fixed point is  guaranteed due to the fact that symmetry-enhanced subspaces of the theoryspace are closed under the RG flow, if the regulator respects the symmetry, just as the regulator does for the global color symmetry  in our case. Thus, the symmetry-enhanced fixed point must also exist in the enlarged theory space. The pivotal question in this context is whether it features more relevant directions. If that were the case, we would have to find a physical interpretation for the distinction of the indices. However it turns out that the fixed point in the enlarged theory space only features additional irrelevant directions, cf.~Tab.~\ref{tab:thetacolorfree}. Thus, the distinction of the couplings by the color structure is a microscopic detail that leaves the universality class intact.

Finally, we mention one additional fixed point that arises in the $(T\bar{T})^4$ truncation and without the distinction of colors in the couplings for the first time. It features four relevant directions,
\bea
\theta_1&=&2.75, \, \theta_{2,3}=2.68 \pm i\, 1.87, \, \theta_4=0.93, \nonumber\\
 \theta_5&=&-0.91, \, \theta_6=-1.02.
\eea
Its coordinates are given by
\bea
g_{4,1}^{2,1}&=& -0.96, \, g_{4,2}^2 = -0.45, \, g_{6,1}^{3,1}=1.67, \, g_{6,1}^{2,1}= -4.42,  \nonumber\\
 g_{6,2}^{3,1}&=& -1.86,\, g_{6,3}^3=11.33,
\eea
and the anomalous dimension is
\be
\eta= -0.998.
\ee
As we do not enlarge the truncation further, it is difficult to say whether this fixed point is merely a truncation artifact. We leave this question for future work.

\section{Conclusions and outlook}

Tensor models are a discrete approach to quantum gravity, related to a sum over triangulations of spacetime. The double-scaling limit, which is a continuum limit taking into account interactions beyond those that triangulate spheres, is linked to the large $N$ limit, where $N$ is the tensor size. Universal scaling behavior in that limit can be related to a Renormalization Group fixed point. 
We have generalized the functional Renormalization Group approach that we have developed for discrete matrix models to the case of rank-3-tensor models. In these models, the notion of Renormalization Group scale is  abstract and not related to a notion of local coarse-graining in position space. Instead,  it relies on the property that the Functional Renormalization Group provides an interpolation between effective descriptions of physics with a decreasing number of degrees of freedom to set up an RG framework. To implement this in practice,
we use the tensor size $N$, which controls the number of degrees of freedom as an RG scale. 

 In particular one can use this RG setup to search for the double-scaling limit, which in this way can be found as a fixed point with one relevant direction and suitable critical exponent. For this purpose, we derive the functional Renormalization Group equation for these models and apply it to a truncation with 16 different couplings including all tensor invariants up to sixth order in the tensors. This truncation already takes into account several distinct combinatorial structures with different geometric interpretation. 
We show how the FRG equation automatically provides a unique assignment of the canonical scaling dimensionality for most of the couplings in the truncation. This allows us to discover interacting fixed points of the RG flow. We re-derive the double scaling limit with a reasonably good approximation of the value of the relevant critical exponent. The 15 additional directions in our truncated theory space are increasingly irrelevant, a posteriori justifying our choice of truncation.
Further, we discover additional fixed points, which we discuss in the context of a possible relation to continuum gravity. These fixed point feature several relevant directions, and could thus be interpreted as underlying a continuum limit beyond double scaling. On the other hand, the relevance of disconnected microscopic interactions (multi-trace-terms) at those fixed points might hint towards a scenario where additional, e.g., matter-like degrees of freedom are present.
We also elucidate how the expectation that certain properties of the microscopic model should not play a role for the universal continuum limit is reflected in the fixed point structure, where the introduction of additional microscopic structure does not lead to additional relevant directions at a fixed point, thus leaving the universality class intact.

As a drawback of the method, the use of the tensor size $N$ as an RG scale necessarily leads to the breaking of the $U(N)\otimes U(N)\otimes U(N)$ symmetry of the model. This is reminiscent of the application of the Renormalization Group to continuum gravity, where the procedure of coarse graining is challenging to reconcile with an intact background independence. In this work, we neglect that the theory space of our model is accordingly enlarged by terms which break the symmetry. We thus set up projection prescriptions onto the couplings which are unique in the symmetric theory space, but which lead to contaminations of our beta functions by contributions from operators with broken symmetry. 

Several distinct but equally critical open questions can be addressed in the future:

Firstly, a systematic extension of the truncation is important to establish the existence and properties of interacting fixed points. In particular, going beyond a finite number of couplings, and exploring, e.g., a truncation of the type ${\rm tr} F(T_{ijk}\bar{T}_{ilm}T_{nlm}\bar{T}_{nop})$ with a function $F$ is now possible. Due to the relevance of the leading term in a Taylor expansion of $F$ at the fixed points that we discovered, we conjecture that this particular truncation could already provide quantitative information on fixed points.

Secondly, the breaking of the $U(N)\otimes U(N)\otimes U(N)$ leads to an enlargement of the theory space which we have ignored. Instead, our projection technique leads to a mixing of symmetric and symmetry-breaking terms. In the future, it will be critical to disentangle these contributions, study the effect of symmetry-breaking operators and explore the consequences of the corresponding modified Ward-identity. 

Thirdly, a model with real tensors and a $O(N)\otimes O(N) \otimes O(N)$ symmetry contains simplicial interactions of the form $T_{ijk}T_{ilm}T_{njm}T_{nlk}$, etc. The model can be explored along the same lines as the model that we have studied here. The possibility of additional interactions with a geometric interpretation makes the study worthwhile.

Finally, our method is straightforward to generalize to rank-4-tensors which provide a discrete description of four-dimensional spacetime. The major difference to the rank-3-model lies in the larger number of invariants already at the quartic level, making the extension to higher order in the tensors slightly more challenging on the technical level. On the other hand, the double-scaling limit should again correspond to a fixed point with only one relevant direction, implying that already small truncations should be sufficient to rediscover the double-scaling limit, and then go beyond to explore further fixed points.

\emph{Acknowledgements}
A.~E.~is supported by the DFG through the Emmy-Noether-program under grant no.~EI-1037-1.

\end{document}